\documentclass[onecolumn,preprintnumbers,amsmath,amssymb]{revtex4}

\begin{document}


\title{Critical points in a relativistic bosonic gas induced by the quantum structure of spacetime}

\author{E. Castellanos}
\email{kruskal@mexico.com}\affiliation{Departamento de F\'{\i}sica,
Universidad Aut\'onoma Metropolitana--Iztapalapa\\
Apartado Postal 55--534, C.P. 09340, M\'exico, D.F., M\'exico.}

\author{Abel Camacho}
\email{acq@xanum.uam.mx} \affiliation{Departamento de F\'{\i}sica,
Universidad Aut\'onoma Metropolitana--Iztapalapa\\
Apartado Postal 55--534, C.P. 09340, M\'exico, D.F., M\'exico.}

\date{\today}

\begin{abstract}
It is well known that phase transitions arise if the interaction
among particles embodies an attractive as well as a repulsive
contribution. In this work it will be shown that the breakdown of
Lorentz symmetry, characterized through a deformation in the
relation dispersion, plus the bosonic statistics predict the
emergence of critical points. In other words, in some quantum
gravity models the structure of spacetime implies the emergence of
critical points even when no interaction among the particle has
been considered.

\end{abstract}

\maketitle
\section{Introduction}
The possibility  that Lorentz symmetry is just an approximation to
quantum space-time can be interpreted, in some cases, as a
deformation in the relation dispersion
\cite{Giovanni1,Giovanni2,Giovanni3,GiovanniPiran}. This
deformation can be characterized, from a phenomenological point of
view, as corrections hinging upon  Planck's length, $\ell_{p}$

\begin{equation}
E^{2}=p^{2}\Bigl[1-\alpha\Bigl(E\ell_{p}\Bigr)^{n}\Bigr]+m^{2}.
\end{equation}

Here $\alpha$ is a coefficient, whose precise value depends upon
the considered quantum gravity model, and $n$, the lowest power in
Planck's length leading to a non-vanishing contribution, is  also
model dependent.

 In ordinary units we have

\begin{equation}
E^{2}=p^{2}c^{2}\Bigl[1-\alpha\Bigl(E\sqrt{G/(c^{5}\,
\hbar)}\Bigr)^{n}\Bigr]+\Bigl(mc^{2}\Bigr)^{2}.
\end{equation}

The most difficult aspect in the search of experimental hints
relevant for the quantum-gravity problem is the smallness of the
involved effects. However, a modified dispersion relation emerges
as an excellent way for searching phenomenological effects in this
kind of theories. In the experimental quest for this kind of
effects, interferometry has played a fundamental role, through the
energy dependence of the speed of light
\cite{Giovanni1,CamachoCastellanos}.

The perspective concerning the development of this type of
theories in quantum gravity appears intimately attached to the
experimental confirmation of some of their predictions.

The idea in the present work is to introduce a deformed dispersion
relation as a fundamental fact for the statistical mechanics of
massive relativistic bosons. Afterwards, we analyze the effects of
this assumption upon the thermodynamics and critical phenomena of
the corresponding gas.

In a physical system comprising a large number of particles either
we consider interaction among the particles or neglect it
\cite{Phatria}. In the first case the existence of phase
transitions and critical phenomena is practically null, with the
exception of the Bose-Einstein condensation, which appears without
the presence of interaction among the components of the system.
The phenomenon of Bose-Einstein condensation is caused exclusively
by quantum effects.

In the subsequent development of this work we will analyze a
relativistic boson gas without including interactions among the
corresponding particles. It will be proved that the bosonic
statistics plus the breakdown of Lorentz symmetry entail the
emergence of critical points. Indeed, we may consider the quantal
properties of the system as an attractive pseudo-interaction,
whereas the breakdown of Lorentz symmetry may be understood as a
repulsive pseudo-interaction. These two ingredients, this is the
main issue in our work, imply the appearance of critical points.
\bigskip
\bigskip

\section{2. DEFORMED RELATION DISPERSION AND BOSONIC STATISTICS}
\bigskip

As mentioned above, several quantum-gravity models suggest a
deformation in the dispersion relation, the one can be
characterized phenomenologically, through corrections hinging upon
Planck's length, $\ell_{p}$ ($E_{p}=c^{5}\hbar/G$ denotes Planck's
energy, where $G$ is the Newtonian gravitational constant, $c$
denotes the speed of light, and $\hbar$ is the Planck's constant
divided by $2 \pi$), and in ordinary units

\begin{equation}
E^{2}=p^{2}c^{2}\Bigl[1-\alpha\Bigl(\frac{E}{E_{p}}\Bigr)^{n}\Bigr]+\Bigl(mc^{2}\Bigr)^{2}.
\end{equation}

Let us consider massive bosons and then the relation between energy
an momentum becomes now:

\begin{equation}
p=\frac{1}{c}\sqrt{\frac{E^{2}-m^{2}c^{4}}{1-\alpha\Bigl(E/E_{p}\Bigr)^{n}}}.
\end{equation}

The number of microstates is given by \cite{Greiner}

\begin{equation}
\Sigma=\frac{s}{\Bigl(2\pi\hbar\Bigr)^{3}}\int\int
d\vec{r}d\vec{p}.
\end{equation}

In this last expression, $s$ is a weight factor arising from the
internal structure of the particles, i.e., spin. If our gas is
inside a container of volume $V$.

\begin{equation}
\Sigma=\frac{4\pi sV}{\Bigl(2\pi\hbar\Bigr)^{3}}\int p^{2}dp.
\end{equation}

And then we obtain the density of states per energy unit

\begin{equation}
\Omega(E)=\frac{4\pi
sV}{(2c\pi\hbar)^{3}}\sqrt{\frac{E^{2}-m^{2}c^{4}}
{1-\alpha\Bigl(E/E_{p}\Bigr)^{n}}}\Bigl\{\frac{E+\alpha[(n-1)E^{2}-nm^{2}c^{4}]
\Bigl(E^{n-1}/E_{p}^{n}\Bigr)}{[1-\alpha(E/E_{p})^{n}]^{2}}\Bigr\}.
\end{equation}

If in (7), we set $\alpha=0$, $s=2$ and $m=0$, then we recover the
density of states for photons \cite{Phatria,Greiner}

\begin{equation}
\Omega(E)=\frac{8\pi V}{(2c\pi\hbar)^{3}}E^{2}.
\end{equation}

Let us analyze the pressure of our gas. In the grand canonical
ensemble for an ideal Bose-Einstein gas \cite{Phatria}, we have

\begin{equation}
\frac{PV}{\kappa T}=-\sum_{E}\ln(1-\lambda e^{-\beta E}).
\end{equation}

Where $\lambda\equiv exp(\mu/\kappa T)$ is the so-called fugacity
and $\kappa$ is Boltzmann's constant.  \cite{Phatria}.

With these assumptions we may obtain that the pressure is given by
\cite{Camacho}

\begin{eqnarray}
P&=&\frac{4\pi s}{(2\pi
c\hbar)^{3}}\Bigl\{\frac{1}{3}\int_{0}^{\infty}\frac{E^{4}}
{\sqrt{E^{2}+m^{2}c^{4}}}\frac{dE}{\lambda^{-1}exp[\sqrt{E^{2}+m^{2}c^{4}/\kappa
T}]-1}\\\nonumber &+&\alpha(n+3/2)(\kappa
T)^{4}(T/T_{p})^{n}\sum_{l=0}^{n}\frac{n!}{l!(n-l)!}
\\\nonumber &\times&\Bigl(\frac{mc^{2}}{\kappa
T}\Bigr)^{l}\Gamma(n+3-l)g_{n+4-l}(\lambda exp(-mc^{2}/\kappa
T))\Bigr\}.
\end{eqnarray}

Where $T_{p}=E_{p}/\kappa$, denotes Planck's temperature,
$g_{\nu}(x)$ are the Bose-Einstein functions \cite{Phatria,Greiner},
and $\Gamma(x)$ are the Gamma functions.

It is important to mention that if $\alpha>0$,  then the pressure
grows, with respect to the case in which Lorentz symmetry is
present. This last remark allows us to interpret the breakdown of
Lorentz symmetry for massive bosons as a repulsive interaction, if
$\alpha>0$. Indeed, the presence of a repulsive interaction (among
the particles of a gas) entails the increase of the pressure,
compared against the corresponding value for an ideal gas. It is
in this sense that we say that the loss of symmetry appears, at
the bulk level, as the emergence of a repulsive interaction (if
$\alpha>0$). The quantum statistics can be contemplated as a
pseudo--attractive interaction, in the sense that a series
expansion in the parameter $N\Lambda^{3}/V$ entails a pressure
lower than that related to an ideal gas \cite{Phatria}. We have
then the main ingredients associated to systems showing phase
transitions.

The Bose-Einstein functions are defined by
 \cite{Phatria,Greiner}

\begin{equation}
g_{\nu}(\lambda)=\frac{1}{\Gamma(\nu)}\int_{0}^{\infty}\frac{x^{\nu-1}dx}{\lambda^{-1}e^{x}-1}=\lambda+\frac{\lambda^{2}}
{2^{\nu}}+\frac{\lambda^{3}}{3^{\nu}}+\,...
\end{equation}

On the other hand \cite{Phatria}

\begin{equation}
g_{3/2}(\lambda)=\frac{\Lambda^{3} N}{V}=\frac{\Lambda^{3}}{v}.
\end{equation}

As a first and roughly approximation:

\begin{equation}
\frac{\lambda^{2}}{2^{3/2}}+\lambda-\frac{\Lambda^{3} N}{V}=0.
\end{equation}

Then we have the fugacity in terms of the volume

\begin{equation}
\lambda=-2^{1/2}+\sqrt{2+2^{3/2}\frac{\Lambda^{3} N}{V}}.
\end{equation}

Where $\Lambda=h/(2\pi m \kappa T)^{1/2}$ is the mean thermal
wavelength \cite{Phatria}. We take the case in which $n=2$ then
the pressure is given by

\begin{eqnarray}
P=\frac{4 \pi s}{(2 \pi c\hbar)^{3}}\Bigl\{\lambda
e^{-mc^{2}/\kappa T}\Bigl[\frac{1}{3}f(T)+
\alpha\Bigl(g(T)+h(T)\lambda e^{-mc^{2}/\kappa T}+j(T)\lambda^{2}
e^{-2mc^{2}/\kappa T}\Bigr)\Bigr]\Bigr\}.
\end{eqnarray}

Where $f(T)$,  $g(T)$,  $h(T)$  and  $j(T)$, are functions of the
temperature, given by

\begin{equation}
f(T)=6(\kappa T)^{4}+6(mc^{2})(\kappa T)^{3}+3(mc^{2})^{2}(\kappa
T)^{2}+(mc^{2})^{3}(\kappa T).
\end{equation}

\begin{equation}
g(T)=\frac{7}{2}(\kappa
T)^{4}\Bigl(\frac{T}{T_{p}}\Bigr)^{2}\Bigl\{\Gamma(5)+2\Gamma(4)\Bigl(\frac{mc^{2}}{\kappa
T}\Bigr)+\Gamma(3)\Bigl(\frac{mc^{2}}{\kappa T}\Bigr)^{2}\Bigr\}.
\end{equation}

\begin{equation}
h(T)=\frac{7}{2}(\kappa
T)^{4}\Bigl(\frac{T}{T_{p}}\Bigr)^{2}\Bigl\{\frac{\Gamma(5)}{2^{6}}+\frac{\Gamma(4)}{2^{4}}\Bigl(\frac{mc^{2}}{\kappa
T}\Bigr)+\frac{\Gamma(3)}{2^{4}}\Bigl(\frac{mc^{2}}{\kappa
T}\Bigr)^{2}\Bigr\}.
\end{equation}

\begin{equation}
j(T)=\frac{7}{2}(\kappa
T)^{4}\Bigl(\frac{T}{T_{p}}\Bigr)^{2}\Bigl\{\frac{\Gamma(5)}{3^{6}}+\frac{2\Gamma(4)}{3^{5}}\Bigl(\frac{mc^{2}}{\kappa
T}\Bigr)+\frac{\Gamma(3)}{3^{4}}\Bigl(\frac{mc^{2}}{\kappa
T}\Bigr)^{2}\Bigr\}.
\end{equation}

If we set $\alpha=0$ , $m=0$, and $\lambda=1$ we recover the
expression for the pressure of a gas of photons \cite{Phatria}

\begin{equation}
P\sim(\kappa T)^{4}.
\end{equation}

Let us expand the fugacity (14) in terms of $(\Lambda^{3}N/V)$
then, pressure is given by

\begin{eqnarray}
P=\frac{4 \pi s}{(2 \pi
c\hbar)^{3}}\Bigl[F(T)\Bigl(\frac{\Lambda^{3}N}{V}\Bigr)+G(T)\Bigl(\frac{\Lambda^{3}N}{V}\Bigr)^2+H(T)
\Bigl(\frac{\Lambda^{3}N}{V}\Bigr)^{3}+J(T)\Bigl(\frac{\Lambda^{3}N}{V}\Bigr)^{4}+...
\Bigr]
\end{eqnarray}

Where:

\begin{equation}
F(T)=\frac{1}{3}f(T)e^{-mc^{2}/\kappa T}+\alpha e^{-mc^{2}/\kappa
T}g(T)+...\hspace{5.45cm},
\end{equation}

\begin{equation}
G(T)=-\frac{2^{3/2}}{3}f(T)e^{-mc^{2}/\kappa
T}+\alpha\Bigl(h(T)e^{-2mc^{2}/\kappa
T}-2^{3/2}g(T)e^{-mc^{2}/\kappa T}\Bigr)+...,
\end{equation}

\begin{eqnarray}\nonumber
\hspace{0.2cm}H(T)&=&\frac{2^{-2}}{3}f(T)e^{-mc^{2}/\kappa T}\\
&+&\alpha\Bigl(j(T)e^{-3mc^{2}/\kappa
T}-2^{5/2}h(T)e^{-2mc^{2}/\kappa T}+2^{-2}g(T)e^{-mc^{2}/\kappa
T}\Bigr)+...,\hspace{1cm}
\end{eqnarray}

\begin{eqnarray}
J(T)=\alpha\Bigl((-2^{3/2}+2^{-1})h(T)e^{-2mc^{2}/\kappa
T}-(2^{5/2}+2^{3/2})j(T)e^{-3mc^{2}/\kappa
T}\Bigr)+...\hspace{0.45cm}
\end{eqnarray}

At the critical points the first and the second derivatives for
the pressure respect to the volume must vanish \cite{GarcíaColín1}

\begin{equation}
\Bigl(\frac{\partial P}{\partial
V}\Bigr)_{T}=0\hspace{2cm}\Bigl(\frac{\partial^{2} P}{\partial
V^{2}}\Bigr)_{T}=0.
\end{equation}

Using this fact we can obtain an expression for the critical
 volume in terms of the temperature
and our additional parameter $\alpha$

\begin{equation}
\frac{V_{c}}{N}=v_{c}=\frac{1}{2}\Lambda^{3}\Bigl\{-3\frac{H(T)}{G(T)}
+\sqrt{9\Bigl(\frac{H(T)}{G(T)}\Bigr)^{2}-24\Bigl(\frac{J
(T)}{G(T)}\Bigr) }\,\,\,\,\,\,\Bigr\}.
\end{equation}

Let us analyze the expression (27), fixing $\alpha=0$. If we
introduce this condition, the critical volume goes to zero, and
the critical pressure goes to infinite. In other words, we recover
the ideal behavior \cite{GarcíaColín2}. This fact allows us to
interpret the parameter $\alpha$ as the intensity of a repulsive
interaction among the particles of our bosonic gas. Let us remark
that this interpretation is related to the quantum structure of
spacetime, and not properly with a real interaction among the
particles of the system.

Let us analyze the critical volume with the experimental values,
$T=50\times10^{-9}\,\ K$ and the mass of $Rb_{87}$ \cite{Pethick},
then (27) becomes
\begin{equation}
v_{c}=\frac{V_{c}}{N}\approx\alpha 10^{-57}m^{3}.
\end{equation}

Due to the limitations in the calculation of the critical
temperature, in terms of the parameter $\alpha$, let us impose a
limit on the experimental parameters. We now analyze the behavior
of the critical volume in the region where the quantum effects are
predominant. In other words, we analyze the behavior of the
critical volume when $T\rightarrow0$. The critical volume, when
$T\rightarrow0$, is given by

\begin{equation}
v_{c}=\frac{V_{c}}{N}=\frac{1}{2}\Lambda^{3}\Bigl[3\eta-\sqrt{9\eta^{2}-24\alpha
|\Theta(T)| }\,\,\,\,\Bigr].
\end{equation}

Where

\begin{equation}
\Theta(T)=\frac{J(T)}{G(T)}=\frac{(-2^{3/2}+2^{-1})h(T)}{-\frac{2^{3/2}}{3}f(T)}e^{-mc^{2}/\kappa
T}
> 0.
\end{equation}

If we set in (29) $\alpha=0$ we recover the ideal behavior.

\section{Conclusions}
\bigskip

Accepting the breakdown of Lorentz symmetry as a fundamental fact
for the statistics of a massive boson gas, we observe that the
pressure of a massive boson gas grows respect to the usual case,
if we assume that $\alpha>0$. This fact allows us interpret the
breakdown of Lorentz symmetry as a repulsive pseudo-interaction.
It has been proved that under these conditions phase transitions
shall emerge. Unfortunately, the critical parameters are very
complicated functions of the temperature, and therefore we do not
obtain an analytic expression for the critical temperature and the
critical pressure in terms of the parameter $\alpha$. In other
words, some quantum gravity models predict the emerge of critical
points, due to the breakdown of Lorentz symmetry in the form of a
modified dispersion relation, even for systems in which no
interaction among the particles exist.

Clearly, a realistic system does include interaction among its
particles, therefore the present model shall be improved
introducing this fact. Additionally, according to the laws of
thermodynamics, the critical exponents, related to phase
transitions, are not independent from one another
 \cite{Phatria,GarcíaColín1,GarcíaColín2}. Therefore we may
 wonder, in this context, if the corresponding exponents arising
 from this scheme satisfy this condition, and what kind of information,
 concerning quantum gravity, could be
 elicited from these exponents.

\bigskip

\begin{acknowledgments}
 This research was partially supported by CONACyT grant 47000--F
and by the México-Germany collaboration grant CONACyT--DFG
J110.491/2006.
\end{acknowledgments}

\end{document}